\documentclass[utf8]{frontiersinFPHY_FAMS} 

\setcitestyle{square}
\usepackage{datetime, fmtcount, etoolbox, fcprefix}
\usepackage{url,hyperref,lineno,microtype,graphicx}
\usepackage[onehalfspacing]{setspace}
\usepackage{tabularx}
\usepackage{float}

\def\keyFont{\fontsize{8}{11}\helveticabold }
\def\firstAuthorLast{Joshi {et~al.}} 
\def\Authors{Yogesh C. Joshi\,$^{1,*}$, Deepak\,$^{1}$, Sagar Malhotra\,$^{2}$}
\def\Address{$^{1}$ Aryabhatta Research Institute of Observational Sciences, Nainital, UK 263002, India\\
$^{2}$ Indian Institute of Science Education \& Research, Mohali 140306, India}
\def\corrAuthor{Yogesh C. Joshi}
\def\corrEmail{yogesh@aries.res.in}
\usepackage{booktabs}

\begin{document}
\onecolumn
\firstpage{1}

\title[Metallicity Gradients in the Galactic Disk]{On the Metallicity Gradients in the Galactic Disk using Open Clusters} 

\author[\firstAuthorLast ]{\Authors} 
\address{} 
\correspondance{} 

\extraAuth{}

\maketitle
\begin{abstract}
    We study the metallicity distribution and evolution in the Galactic disk based on the largest sample of open star clusters in the Galaxy. From the catalogue of 1879 open clusters in the range of Galactocentric distance ($R_{\rm GC}$) from 4 to 20 kpc, we investigate the variation of metallicity in the Galactic disk as functions of $R_{\rm GC}$, vertical distance ($Z$), and ages of the clusters. In the direction perpendicular to the Galactic plane, variation in metallicity is found to follow a stepped linear relation. We estimate a vertical metallicity gradient $\frac{d{\rm [Fe/H]}}{d|Z|}$ of $-0.545\pm0.046$ dex kpc$^{-1}$ for $|Z| < 0.487$ kpc, and $-0.075 \pm 0.093$ dex kpc$^{-1}$ for $0.487 < |Z| < 1.8$ kpc. On average, metallicity variations above and below the Galactic plane are found to change at similar rates. The change in metallicity in the radial direction is also found to follow a two-function linear relation. We obtain a radial metallicity gradient $\frac{d{\rm [Fe/H]}}{dR_{\rm GC}}$ of $-0.070\pm0.002$ dex kpc$^{-1}$ for $4.0\lesssim R_{\rm GC} \lesssim 12.8$ kpc, and $-0.005\pm 0.018$ dex kpc$^{-1}$ for $12.8 \lesssim R_{\rm GC} \lesssim 20.5$ kpc which clearly shows a strong variation in the metallicity gradient when moving from the inner to the outer Galactic disk. Age-metallicity relation (AMR) is found to follow a steeper negative slope of $-0.031\pm0.006$ dex Gyr$^{-1}$ for clusters older than 240 Myr, however, there is some hint of positive metallicity age gradient for younger clusters.
\section{}
\tiny
 \keyFont{\section{Keywords:} Galaxy, Open clusters, Metallicity distribution, Metallicity abundance gradients, Age-metallicity relation}
\end{abstract}

\section*{Introduction}\label{sec-intro}
For a long time, open clusters (OCs) have been used to trace the Kinematical, dynamical and chemical evolution of the Galaxy (\cite{AllenCarigiPeimbert1998ApJ...494..247A, 2014A&A...572A..92M, 2019AstL...45..109B}). Since OCs span a wide range of ages and chemical compositions and mostly lie in the Galactic plane, they are identified as tracers of the Galactic disk (\cite{2011AJ....142...51L, 2014ApJ...788...89T, 2023AJ....166..170J}). As ages and chemical compositions of OCs can be determined with higher precision in comparison to the field stars, they are believed to be better tracers of temporal and chemical evolution of the Galactic properties (\cite{NetopilPaunzenHeiter2016AA...585A.150N, 2017A&A...603A...2M, DonorFrinchaboyCunha2018AJ....156..142D, 2021ApJ...919...52Z, SpinaTingDeSilva2021MNRAS.503.3279S, NetopilOralhanInc2022MNRAS.509..421N}). With over 6000 OCs discovered in the Galaxy so far, we now have a much better understanding of their properties and, as a result, of the composition of the Galaxy (\cite{2023AJ....166..170J, MagriniViscasillasVazquezSpina2023AA...669A.119M}). In recent years, the number of OCs having metallicity information has increased significantly, with the large-scale spectroscopic surveys such as the Gaia-ESO Public Spectroscopic Survey (\cite{2017A&A...603A...2M}), GALAH (\cite{2017MNRAS.465.3203M}), APOGEE (\cite{2017AJ....154...94M, DonorFrinchaboyCunha2020AJ....159..199D}) and the LAMOST survey (\cite{ZhongChenWu2020AA...640A.127Z}). Furthermore, due to the availability of high-quality photometric and astrometric data from the ESA Gaia mission (\cite{2016A&A...595A...1G}), significant improvement has been made in the ability to refine the cluster membership resulting in a better estimate of age and distance among other parameters (\cite[e.g.,][]{2018A&A...618A..93C, 2020A&A...640A...1C, DiasMonteiroMoitinho2021MNRAS.504..356D}).

Radial abundance gradient is one of the key constraints to the Galactic chemical evolution models. The exact nature of the radial metallicity gradient, reported through various tracers like Planetary Nebulae, HII region, OB stars, Classical Cepheids, etc., is still not quite conclusive and portrays a diverse picture of the chemical evolution of the Galaxy (\cite[e.g.,][and references therein]{AndrievskyKovtyukhLuck2002A&A...381...32A, DaflonCunha2004ApJ...617.1115D, MacielLagoCosta2005A&A...433..127M, LemasleFrancoisPiersimoni2008A&A...490..613L, 2010A&A...512A..19M, GenovaliLemasleBono2014AA...566A..37G, 2023A&A...678A.195D}). However, having an extensive range in age, distance, and chemical composition, OCs are regarded as a better tracer than other such sources (\cite{2008IAUS..248..433C, 2013pss5.book..347F, MagriniViscasillasVazquezSpina2023AA...669A.119M}). Various studies have been carried out in the last two decades to study the chemical evolution of the Galactic disk using OCs (e.g., \cite{FrielJanesTavarez2002AJ....124.2693F,  ChenHouWang2003AJ....125.1397C, BragagliaSestitoVillanova2008AA...480...79B, 2010AJ....139.1942F, 2011A&A...535A..30C, YongCarneyFriel2012AJ....144...95Y, 2013ApJ...777L...1F, ReddyLambertGiridhar2016MNRAS.463.4366R, NetopilPaunzenHeiter2016AA...585A.150N, 2017A&A...603A...2M}. However, the main advancement came after the recent release of three large-scale surveys namely Gaia-ESO (\cite{RandichGilmoreMagrini2022AA...666A.121R}), GALactic Archeology with HERMES (GALAH; \cite{2017MNRAS.465.3203M}), and Apache Point Observatory Galactic Evolution Experiment (APOGEE; \cite{2017AJ....154...94M}) which resulted in estimation of more complete and precise chemical compositions of a large number of OCs (\cite[e.g.,][]{CarreraBragagliaCantat-Gaudin2019AA...623A..80C, ZhongChenWu2020AA...640A.127Z, DonorFrinchaboyCunha2020AJ....159..199D, 2021ApJ...919...52Z, SpinaTingDeSilva2021MNRAS.503.3279S, 2022Univ....8...87S,  NetopilOralhanInc2022MNRAS.509..421N, 2022AJ....164...85M, MagriniViscasillasVazquezSpina2023AA...669A.119M}). These studies have obtained a single slope radial metallicity gradient ranging from -0.051 to -0.077 dex/kpc while employing a two-function radial metallicity gradient, they obtained a steeper slope in the range of -0.054 to -0.081 dex/kpc for the younger and inner region of clusters while a shallower slope of 0.009 to 0.044 dex/kpc for the older and outer region of clusters which is also supported by inside-out disk formation models. The intersection point or knee point in such two-function slopes also varies between 11 to 12 kpc among different studies. The age-metallicity relation is one other important constraint on the theoretical models of the Galactic disk and has been studied by various authors using different stellar populations (\cite[e.g.,][]{1995ARA&A..33..381F, CarraroNgPortinari1998MNRAS.296.1045C, 2001A&A...377..911F}). Except for few studies like \cite{1993A&A...275..101E, ChenHouWang2003AJ....125.1397C, ZhongChenWu2020AA...640A.127Z}, most of the studies found no obvious AMR for the OCs population (\cite{2010AJ....139.1942F, 2012AJ....144...95Y, 2021ApJ...919...52Z, NetopilPaunzenHeiter2016AA...585A.150N, 2017A&A...603A...2M}).

Considering a wide range of Galactic chemical evolution parameters among different studies, the prime motive of the present work is to form a more extensive set of OCs having chemical compositions available through recent photometric and spectroscopic surveys, thus extending the sample with a wide range in the age and galactocentric distance. Despite extracting cluster parameters from different sources hence making a heterogenous data set, we trust a statistical analysis on a larger sample of OCs would not lead to any systematic bias in our results. This paper is structured as follows: we describe the data used in the present work in Section~\ref{s_data}. The metallicity distribution of OCs is analyzed in Section~\ref{sec:MetallicityFunctions}. The cluster age-metallicity relation is examined in Section~\ref{sec:FeH_logAge}. In Section~\ref{sec:Fe_Gradient_Rad_Vert}, we investigate various correlations between radial and vertical metallicity gradients with the age and positions of the OCs. Our results are summarized in Section~\ref{sec:discuss}.
\section{Data}\label{s_data}
To understand the chemical evolution of the Galaxy, particularly the Galactic disk, over the last few billion years, a large and homogeneous sample of OCs with measured metallicity and age is required. For this purpose, we searched the literature for OCs with known metallicity along with other information like position coordinates, radial distances, age, etc. It may be noted here that we have used the term metallicity for the iron abundance [Fe/H] (relative to the solar abundance) throughout this study. Most of the OC metallicity estimates reported prior to 2018 are either based on the photometric techniques (\cite{KharchenkoPiskunovSchilbach2013A&A...558A..53K}) or low-resolution spectroscopic data (\cite[e.g.,][and references therein]{NetopilPaunzenHeiter2016AA...585A.150N}). Additionally, over the years, many of the clusters have been studied repeatedly, and so metallicity estimates for these clusters are available based on different techniques, spectral resolutions, and data qualities. To create a comprehensive list of OCs with the best available metallicity estimate, we started by collecting all the metallicity estimates along with other related information like method of estimation (photometric or spectroscopic), spectral resolution, signal-to-noise ratio (SNR), number of member stars used for average metallicity estimation, and the year of reporting from all the major studies published in the last three decades. This resulted in a total of 4772 metallicity reporting for known OCs from \cite{BaratellaDOraziCarraro2020AA...634A..34B, BragagliaSestitoVillanova2008AA...480...79B, CaetanoDiasLepine2015NewA...38...31C, CarraroBresolinVillanova2004AJ....128.1676C, CarraroGeislerVillanova2007AA...476..217C, CarraroMoitinhoZoccali2007AJ....133.1058C, CarraroVillanovaDemarque2008MNRAS.386.1625C, Carrera2012AA...544A.109C, CarreraCasamiquelaOspina2015AA...578A..27C, CarreraBragagliaCantat-Gaudin2019AA...623A..80C, CasamiquelaBlanco-CuaresmaCarrera2019MNRAS.490.1821C, CasamiquelaSoubiranJofre2021AA...652A..25C, ClariaLapassetMinniti1989AAS...78..363C, ClariaPiattiLapasset2003AA...399..543C, ClariaPiattiMermilliod2008AN....329..609C, ConradScholzKharchenko2014AA...562A..54C, DOraziRandich2009AA...501..553D, DeSilvaFreemanAsplund2007AJ....133.1161D, DeSilvaCarraroDOrazi2015MNRAS.453..106D, DiasMonteiroMoitinho2021MNRAS.504..356D, DonatiCocozzaBragaglia2015MNRAS.446.1411D, DonorFrinchaboyCunha2018AJ....156..142D, DonorFrinchaboyCunha2020AJ....159..199D, FordJeffriesSmalley2005MNRAS.364..272F, FossatiFolsomBagnulo2011MNRAS.413.1132F, FrielBoesgaard1992ApJ...387..170F, FrielJanesTavarez2002AJ....124.2693F, FrinchaboyMunozMajewski2004astro.ph.11127F, FrinchaboyThompsonJackson2013ApJ...777L...1F, FuBragagliaLiu2022AA...668A...4F, GeislerVillanovaCarraro2012ApJ...756L..40G, GonzalezWallerstein2000AJ....119.1839G, GrattonContarini1994AA...283..911G, HasegawaSakamotoMalasan2008PASJ...60.1267H, HillPasquini1999AA...348L..21H, JacobsonFrielPilachowski2008AJ....135.2341J, JacobsonFriel2013AJ....145..107J, KrisciunasMonteiroDias2015PASP..127...31K, Luck1994ApJS...91..309L, MagriniRandichZoccali2010AA...523A..11M, MagriniSpinaRandich2018AA...617A.106M, MargheimKingDeliyannis2000AAS...196.4210M, MonroePilachowski2010AJ....140.2109M, MyersDonorSpoo2022AJ....164...85M, NetopilPaunzen2013AA...557A..10N, NetopilPaunzenHeiter2016AA...585A.150N, NetopilOralhanInc2022MNRAS.509..421N, OverbeekFrielJacobson2016ApJ...824...75O, PasquiniRandichZoccali2004AA...424..951P, PaunzenMaitzenRakos2003AA...403..937P, PaunzenHeiterNetopil2010AA...517A..32P, PereiraQuireza2010IAUS..266..495P, PiattiClariaAbadi1995AJ....110.2813P, RandichGilmoreMagrini2022AA...666A.121R, ReddyGiridharLambert2013MNRAS.431.3338R, SantosLovisPace2009AA...493..309S, SantosLovisMelendez2012AA...538A.151S, SchulerKingFischer2003AJ....125.2085S, SestitoRandichMermilliod2003AA...407..289S, SpinaTingDeSilva2021MNRAS.503.3279S, TwarogAshmanAnthony-Twarog1997AJ....114.2556T, VanseviciusPlataisPaupers1997MNRAS.285..871V, VillanovaCarraroBresolin2005AJ....130..652V, WarrenCole2009MNRAS.393..272W, YongCarneyFriel2012AJ....144...95Y, ZacsAlksnisBarzdis2011MNRAS.417..649Z, ZhongChenWu2020AA...640A.127Z} and reference therein, which also includes multiple reporting from different studies for some clusters. To avoid duplication of OCs in the list because of the use of different identifiers for a cluster in different studies, we made use of the {\tt astroquery.simbad} package to detect and assign a common name to all such duplicates. As a secondary measure, we manually searched and checked all the possible duplicates with a spatial angular distance of less than 0.1 degrees along with a maximum difference of 5 milli-arcsec in OC's proper motion. This helped us to detect and eliminate five more duplicate OCs pairs: Berkeley 85 -- Dolidze 41, COIN-Gaia23 -- Majaess 65, NGC 1746 -- NGC 1750, vdBergh-Hagen 72 -- UBC 491, and vdBergh-Hagen 84 -- Gulliver 35. Some of the other OCs pairs, like UBC 55 -- FSR 686 and UBC 73 -- Gulliver 56, have very small separations in phase space but are confirmed different clusters in previous studies. For example, \cite{PieckaPaunzen2021A&A...649A..54P} suggests that UBC 55 and FSR 686 are a possible pair of binary clusters while UBC 73 and Gulliver 56 are also different clusters.

To select the best unique metallicity estimate from multiple reporting for each of the clusters, we selected metallicity estimates by giving higher priority to spectroscopic studies (compared to the photometric metallicity estimate), followed by highest spectral resolution, highest SNR, highest number of member stars used to find the average metallicity for the OC, smallest error in the reported metallicity, and latest reporting. This resulted in a final sample of 1879 unique OCs with the best available metallicity estimates, of which 615 have metallicity estimates based on spectroscopic data (hereafter sample OCS, where ``S'' stands for spectroscopic) and the remaining 1264 have metallicity estimates based on the photometric data (hereafter sample OCP, where ``P'' stands for photometric).

Like metallicity selection, we selected the best quality astrometric and age data for each cluster from multiple reportings by prioritizing the most recent publication. During the selection, preference was given to studies that also provide measurement uncertainties. From our final sample of 1879 unique clusters, astrometric data (including distance information) is available for all the clusters while age is available for all but one cluster. However, uncertainty estimates in age and metallicity parameters for all the clusters could not be found therefore, any weighted statistical analysis cannot be done in the present study.

Adopting the Galactocentric distance of the Sun, $\rm R_\odot$, as 8.15 kpc (\cite{ReidMentenBrunthaler2019ApJ...885..131R}), we calculated the Galactocentric distance of the cluster using the following well-known transformation relation.
\begin{equation}
R_{\rm GC} = \sqrt{{\rm R_\odot^{2}}+ (d~cos(b))^{2}-2 {\rm R_\odot}~d~cos(l)~cos(b)}
\end{equation}
where $d$, $l$, and $b$ are the heliocentric distance, Galactic longitude, and Galactic latitude, respectively. We also used the rectangular coordinate system ($X$, $Y$, $Z$), which is defined as $X = d~cos(b)~cos(l)$; $Y = d~cos(b)~sin(l)$; and $Z = d~sin(b)$. The most distant cluster with metallicity information is at $R_{\rm GC} = 20.38$ kpc and only nine OCs are seen beyond $R_{\rm GC} = 14$ kpc. This reveals either a lack of OCs in the outer Galactic disk or observational limitations to observe such clusters due to large extinction along the line of sight. Additionally, the number of OCs decreases drastically as we move farther away from the heliocentre. For example, only 22 OCs are located beyond a heliocentric distance of 5 kpc, further suggesting that the drop in OCs number with radial distance is primarily linked to the detection limits (\cite[e.g.,][]{2016A&A...593A.116J}).

\section{Metallicity distributions in open clusters}\label{sec:MetallicityFunctions}
The metallicity in our cluster sample ranges from about -0.80 to 0.60 dex except for six OCs, NGC 6204, NGC 2129, Trumpler 33, Dolidze 5, NGC 6910 and FSR 932, for which adopted [Fe/H] based on our selection criteria are -1.05, -1.53, -1.54, -1.94, -1.96, and -2.17, respectively. For all the six clusters, adopted metallicities are based on spectroscopic data. For NGC 6204 and Trumpler 33, metallicity estimates are adopted from \cite{ConradScholzKharchenko2014AA...562A..54C} which provided the metallicity estimates based on a spectral resolution of 7500, while for NGC 2129, Dolidze 5, NGC 6910 and FSR 932, metallicity estimates are adopted from \cite{FuBragagliaLiu2022AA...668A...4F} which provided metallicity estimate based on data from LAMOST survey with a spectral resolution of 1800. For NGC 2129 and NGC 6910, \cite{ZhongChenWu2020AA...640A.127Z} provided independent spectroscopic metallicity estimates of $-1.426\pm0.856$ and $-1.97$, respectively, based on data from LAMOST survey with a spectral resolution of 1800. For all of these six clusters, NGC 6204, NGC 2129, Trumpler 33, Dolidze 5, NGC 6910, and FSR 932, \cite{DiasMonteiroMoitinho2021MNRAS.504..356D} provided independent photometric metallicity estimates of  $0.096\pm 0.004$, $-0.07\pm0.01$, $0.145\pm 0.016$, $-0.033 \pm 0.033$,  $0.035\pm0.008$ and $-0.142 \pm 0.008$, respectively. For NGC 6204, \cite{NetopilPaunzenHeiter2016AA...585A.150N} and \cite{PaunzenHeiterNetopil2010AA...517A..32P} also provided photometric metallicity of 0.02 and $-0.14\pm0.10$, respectively. The wrong identification of cluster member stars to estimate the cluster's average metallicity appears to be one of the main reasons for the large differences between the available spectroscopic and photometric metallicities for these six clusters. Finding the exact reason for this discrepancy is beyond the scope of this study. However, considering the unexpectedly lower metallicity and the large difference when compared to available photometric estimates for these six clusters, we exclude these six clusters from further analysis in this study. The final catalogue of 1879 OCs used in this study is provided in a machine-readable format in Table \ref{table:Data_Table}. Among 1879 clusters, 609 have metallicity estimates based on spectroscopic data (sample OCS), and the remaining 1264 have metallicity estimates based on the photometric data (sample OCP).

\begin{figure}
  \centering
\includegraphics[width=0.49\textwidth]{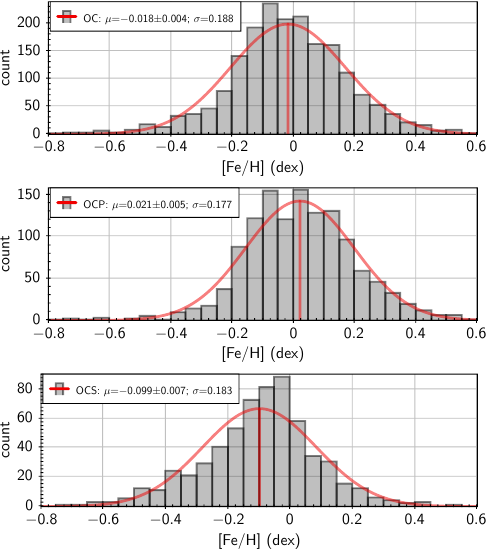}
\caption{Metallicity functions for the sample OC, OCP, and OCS are shown, and a Gaussian fit is applied to all three subsamples with fit parameters provided in the legends.}
\label{fig:FeH_hist_OCs}
\end{figure}

\begin{figure}
  \centering
\includegraphics[width=0.48\textwidth]{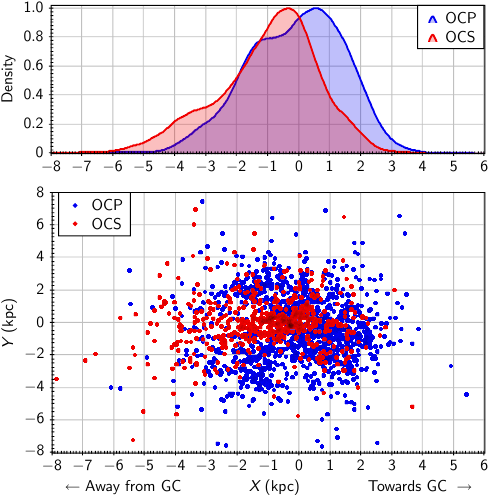}
\caption{Sample OCP and OCS in the {\it X-Y} plane in the Heliocentric frame, where ({\it X, Y}) = (0, 0) corresponds to the location of the Sun and positive {\it X} points towards the Galactic centre. The top panel shows the corresponding density distributions along the {\it X} direction.}
\label{fig:XY_OCP_OCS}
\end{figure}

The metallicity functions for the sample OCP, OCS, and OCs (OCP + OCS) are shown in Figure \ref{fig:FeH_hist_OCs}. For all three cases, the Gaussian distribution fits are also drawn. The OCs sample has a mean metallicity of $-0.018\pm0.004$ with a sample standard deviation ($\sigma$) of 0.188. The sample of OCP has a slightly higher mean metallicity with [Fe/H] = $-0.021 \pm 0.005$ compared to the sample of OCS which has mean [Fe/H] = $-0.099 \pm 0.007$. Both OCP and OCS span an almost similar range in [Fe/H] and also have similar sample standard deviations. The small but significant difference between the mean metallicity of the sample OCP and OCS may be the result of two factors: 1) systematic offsets in the photometric metallicity estimates, and 2) bias in the sample selection. Because of the unavailability of the photometric data for all the cases, it is not possible to directly check for the systematic offsets in estimated metallicities. However, as sample OCP consists of photometric metallicity estimates from many studies that provide metallicity estimates based on different sets of photometric data along with theoretical isochrones, a systematic offset in all or the majority of these studies is not expected. To examine whether bias in the sample selection is the reason behind this offset, we draw the distribution of safmple OCP and OCS in {\it X-Y} plane in the Heliocentric frame in the bottom panel of Figure \ref{fig:XY_OCP_OCS}. Here, the Sun is located at ({\it X, Y}) = (0, 0) where positive {\it X} points towards the Galactic centre (GC) and positive $Y$ points towards the North Galactic Pole. The distribution readily suggests that the OCP clusters are located more towards the GC than the clusters in OCS. This is more clear from the top panel of the figure where the density distributions of the sample OCP and OCS along {\it X} are provided. Here, the distribution for sample OCS is more negatively skewed with a skewness of -0.23 compared to sample OCP which has a skewness of -0.15, and is understood to be due to the presence of more clusters in the anti-GC direction in sample OCS than in the sample OCP.
\section{Age-Metallicity Relation}\label{sec:FeH_logAge}
Age-metallicity relation (AMR) in the Galactic disc is crucial to constrain the chemical evolution models, and star clusters offer an important advantage in the studies of the evolution of the Galaxy because they provide a time sequence for investigating the changes that occur in our Galaxy over the period of time. The large temporal range in the age and metallicity for the OCs offers useful insights related to the chemical evolution history of the Galaxy and also presents a useful constraint on the various theoretical models of the disk (\cite{1995ARA&A..33..381F}). Over the last 20 years, many studies that focus on this relation use either nearby stars (\cite{CarraroNgPortinari1998MNRAS.296.1045C, 2001A&A...377..911F}) or OCs (\cite{NetopilPaunzenHeiter2016AA...585A.150N, 2017A&A...603A...2M,  2023PARep...1...11D}). As noted in many earlier studies, there is no obvious AMR for the OC population (\cite[e.g.,][]{2010AJ....139.1942F, 2012AJ....144...95Y, 2021ApJ...919...52Z}) while some of the studies find a weak AMR (\cite{1993A&A...275..101E, ChenHouWang2003AJ....125.1397C, ZhongChenWu2020AA...640A.127Z}). However, the large sample of OCs having metallicity measurements in the present study is re-employed to understand this relation in some detail.

\begin{figure}
\centering
\includegraphics[width=0.48\textwidth]{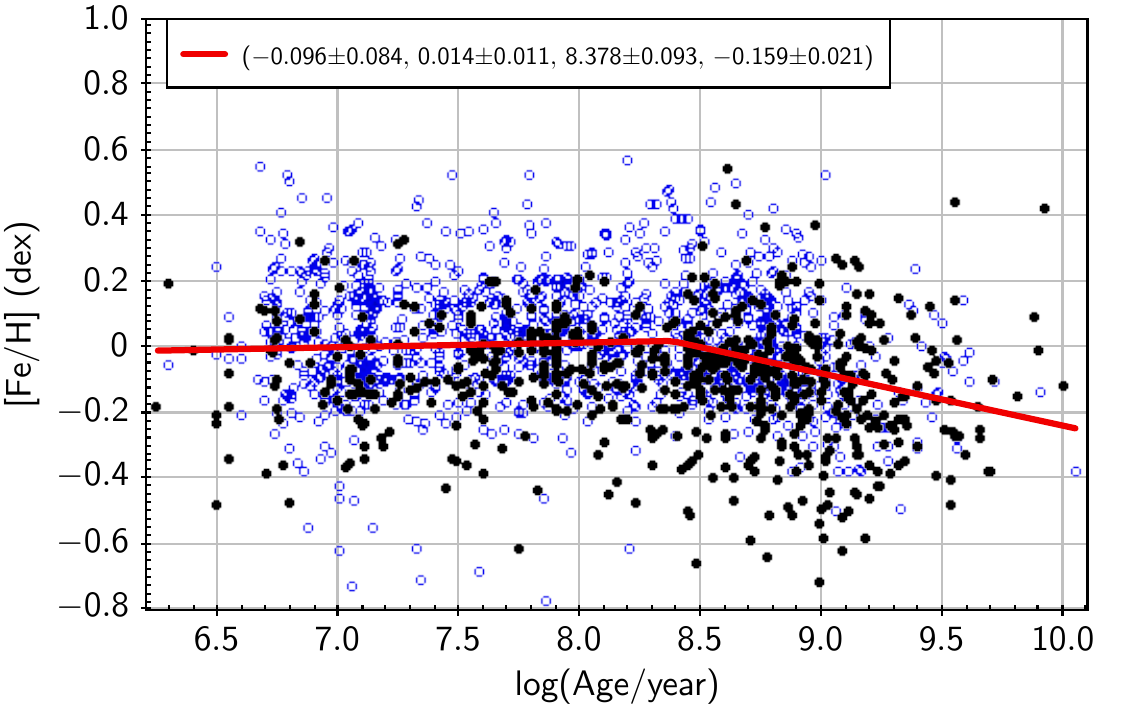}
\caption{Age-metallicity functions for the sample clusters. Open blue circles and filled black circles are the clusters from sample OCP and OCS, respectively. The red line shows the fitted stepped linear regression to the total sample, coefficients (along with corresponding standard errors) for which are provided in the legends in the form (b$_1$, m$_1$, C, m$_2$), where b$_1$ and m$_1$ are the y-axis intercept and slope for the first linear function, m$_2$ is the slope for the second linear function, and C is the point where these two function intersects.
}
\label{fig:FeH_Age}
\end{figure}

The AMR in our sample of clusters is shown in Figure \ref{fig:FeH_Age}. The distribution readily suggests that for clusters with $\log(\rm age/yr) \lessapprox 8.4$, the average metallicity of clusters is near to solar and does not change significantly over time. However, for $\log(\rm age/yr) \gtrapprox 8.4$, clusters average metallicity follows a slightly decreasing trend with an increase in $\log(\rm age/yr)$. To find the exact age-metallicity gradients and the age turn-off point at which the metallicity gradient changes, we fitted the data with a combination of two linear regressions (i.e., stepped linear regression) in the following form: 
\begin{equation}
 {\rm [Fe/H]}  = {\rm m_1} \times \log{(\rm age/yr)} + {\rm b_1},  ~~~~ \log{(\rm age/yr) \leq C}
\label{eq:dFe_logAge_1_form}
\end{equation}
\begin{equation}
 {\rm [Fe/H]}  = {\rm m_2} \times \log{(\rm age/yr)} + {\rm b_2},  ~~~~ \log{(\rm age/yr) > C}
\label{eq:dFe_logAge_2_form}
\end{equation}
where, C is the point of intersection, b$_1$ and b$_2$ are the [Fe/H]-axis intercepts, and m$_1$ and m$_2$ are slopes for the two functions. The coefficients of the fitted regressions (along with the point of intersection C and corresponding standard errors) were determined using iterative least square estimation by treating b$_1$, m$_1$, C, and m$_2$ as variables while assuming b$_2$ = (m$_1 \times \rm C + b_1$). Based on the distribution in Figure \ref{fig:FeH_Age}, we assumed the initial values of  b$_1$, m$_1$, C, and m$_2$ as 0, 0, 8.5, and -0.02, respectively. Estimated values of coefficients from each iteration were adopted as inputs for the next iteration. Along with minimizing the mean error, we also repeated the iterations until the difference between the estimated coefficients and the corresponding adopted coefficients was less than $10^{-5}$. In addition, for each of the iterations, we also checked the distribution of fitted metallicity residual (i.e., observed metallicity value minus predicted model value) as a function of $\log(\rm age/yr)$ and found that for the final iteration, the slope and the intercept to this distribution were less than $10^{-5}$. The final fitted function to our OCs data is shown as the red line in Figure \ref{fig:FeH_Age}. The coefficients of the fitted functions are also provided in the legend of Figure~\ref{fig:FeH_Age} in the form (b$_1$, m$_1$, C and m$_2$). Based on the sample OC, the two linear functions intersect at $\log (\rm age/yr) = 8.378 \pm 0.093$ (i.e., an age of about 240 Myr), and age-metallicity gradients are given as:
\begin{equation}
  \begin{array}{l}
    \frac{d{\rm [Fe/H]}}{d {\log \rm (age/yr)}} = 0.014 \pm 0.011,\\ ~~~~~~~~~~~~~~~~~~~~~~~~~ \log (\rm age/yr) \leq 8.378 \pm 0.093
\label{eq:dFe_logAge_1}
  \end{array}
\end{equation}
\begin{equation}
  \begin{array}{l}
    \frac{d{\rm [Fe/H]}}{d {\log \rm (age/yr)}} = -0.159 \pm 0.021,\\ ~~~~~~~~~~~~~~~~~~~~~~~~~ \log (\rm age/yr) > 8.378 \pm 0.093
\label{eq:dFe_logAge_2}
  \end{array}
\end{equation}

For $\log (\rm age/yr) > 8.378 \pm 0.093$, the decrease in [Fe/H] with an increase in $\log(\rm age/yr)$ with a slope of $\frac{d{\rm [Fe/H]}}{d {\log \rm (age/yr)}} = -0.159 \pm 0.021$ that is equivalent to $-0.031\pm0.006$ dex/Gyr. The negative slope between age and metallicity suggests that the metallicity in the interstellar medium of the Galaxy gradually increased with time until about 240 Myrs ago. In Table \ref{table:Age_FeH_slope}, we compare our derived AMR slope with earlier studies that were carried out using OCs, although with a significantly smaller sample (\cite{2010A&A...511A..56P, ZhongChenWu2020AA...640A.127Z}). Our present estimate is quite consistent with these studies. However, for $\log (\rm age/yr) \leq 8.378 \pm 0.093$, a very slightly increasing trend in [Fe/H] is seen with the increase in $\log(\rm age/yr)$. Though it surprisingly suggests that the formation site of the younger cluster is relatively metal-poor compared to the intermediate age clusters but a slope of 0.014 at almost 1-$\sigma$ level is too small to make any definite conclusion. Overall, a negative slope in AMR is in agreement with the results from previous studies that the metallicity of old-age OCs is lower than that of young and intermediate-age OCs at any given Galactocentric distance (\cite[e.g.,][]{2016A&A...591A..37J, NetopilPaunzenHeiter2016AA...585A.150N, 2017A&A...601A..70S}).  Using a homogenous compilation of 172 clusters from the literature, \cite{NetopilPaunzenHeiter2016AA...585A.150N} investigated the metallicity distribution and found that the clusters younger than 500 Myrs may be characterized by lower metallicities than the older ones, at least in the region between 7 and 9 kpc from the GC. At the same time, they confirmed a negative gradient for these clusters. However, their sample did not include any clusters younger than ~100 Myrs located in the inner Galaxy. 

\begin{figure}
  \centering
\includegraphics[width=0.49\textwidth]{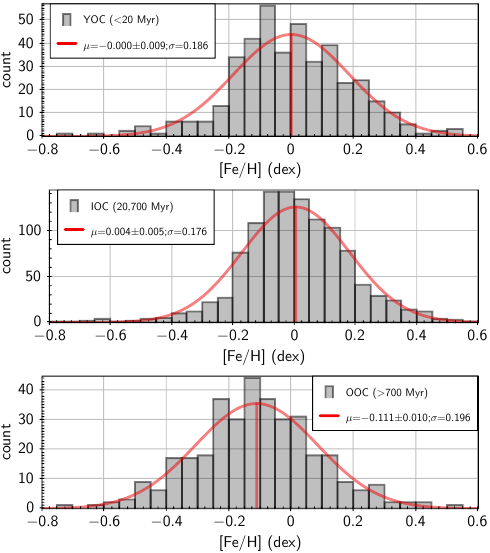}
\caption{The metallicity functions for the samples of YOCs, IOCs, and OOCs are shown along with the Gaussian fit for each population.
}
\label{fig:FeH_hist_YIO_OCs}
\end{figure}

As clearly evident from Figure \ref{fig:FeH_Age}, sample OCS has a relatively lower average metallicity compared to sample OCP at all the ages in the available age span. This, as discussed previously in Section \ref{sec:MetallicityFunctions}, is possibly due to sample selection bias as sample OCP has more clusters from the GC direction while sample OCS has more clusters from the anti-GC direction. More interestingly, both samples have a nearly constant spread in metallicity throughout the available age span suggesting that at both older and recent times the natal gas at the formation site of the clusters had similar mixture properties.
To understand the properties of OCs from different ages, we broadly segregate our sample OC in three different age bins, namely $\le$ 20 Myrs as young open clusters (YOC), 20 to 700 Myr as intermediate-age clusters (IOC), and $>$ 700 Myr as old clusters (OOC). Sample YOC, IOC, and OOC have 410, 1114 and 349 clusters, respectively. Metallicity functions for the sample YOC, IOC and OOC are shown across the panels in Figure \ref{fig:FeH_hist_YIO_OCs}. For YOC, IOC, and OOC, the mean of the [Fe/H] distributions are $-0.000 \pm 0.009$, $0.004 \pm 0.005$, and $-0.111 \pm 0.010$, respectively, and corresponding sample standard deviations are 0.186, 0.176, and 0.196, respectively. There is hardly any significant difference in metallicity between YOCs and IOCs. However, slightly higher mean metallicity for the YOC and IOC compared to OOC is apparent which is well expected as the clusters belonging to YOC and IOC are understood to have formed from gas and dust in the thin disk of the Galaxy that has already been enriched through the earlier generation of stellar formation. Additionally, as readily visible from the figure, the metallicity functions for the OCs of different age groups are not symmetric and are slightly skewed. The skewness for YOC, IOC, and OOC samples are -0.175, -0.212, and 0.083, respectively. All three age group OC samples span almost similar ranges in metallicity. The metal-poor clusters in the YOC sample have likely formed from the fall of a metal-poor gas to the younger thin disk along with the succeeding starburst. This in-fall of metal-poor gas is believed to be due to merging satellite galaxies to the Milky Way (\cite{1999Ap&SS.267..145W}) resulting in diluting the metallicity of interstellar material in the Galactic thin disk and subsequently triggering the formation of a large number of metal-poor clusters. On the other hand, the super solar metallicity clusters in the OOC group may have actually formed from the highly processed material from the inner region of the Galaxy. It is believed that the metal-rich old clusters in the inner region had migrated outwards outer disk over a period of time in order to survive the destruction due to relatively stronger Galactic potential in the inner disc (\cite{2022AJ....164...85M, MagriniViscasillasVazquezSpina2023AA...669A.119M}).

\begin{figure}
\begin{center}
\includegraphics[width=0.49\textwidth]{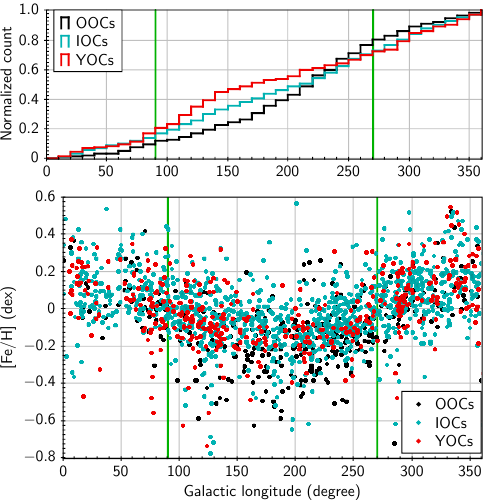}
\caption{
Metallicity as a function of the Galactic longitude for the OCs belonging to the three different age groups. Cumulative distribution along Galactic longitude for OCs in three different age groups is shown in the top panel. In both panels, the clusters between the two vertical green lines (at l = 90$^o$ and 270$^o$) are in the anti-GC direction while the clusters outside these lines are in the GC direction.
}
\label{fig:FeH_Longitude}
\end{center}
\end{figure}

To further understand the reason behind the almost similar large spread in metallicity distribution for OCs of different age groups, we looked into the distribution of [Fe/H] as a function of the Galactic longitude. As shown in the bottom panel of Figure \ref{fig:FeH_Longitude}, clusters towards the anti-GC direction (i.e., with $90^o < l < 270^o$) have relatively lower metallicity than the cluster in the GC direction (i.e. with $270^o < l < 90^o$). The reason behind this asymmetry is that most of the star-forming regions are in the GC direction where the nucleosynthesis process is more active in comparison to lesser present star-forming regions in the anti-GC direction. As a result, metallicity increases as the stellar evolution progresses. The top panel of the figure shows the cumulative distribution functions (CDFs) for the three age groups and readily suggests that all three age populations span a similar range in the Galactic longitude. We further performed the Kolmogorov-Smirnov (KS) test to check if the three CDFs come from the same distribution. The KS test $p$-value for OOC and IOC pair is 0.98, for IOC and YOC pair is 0.87, and for OOC and YOC pair is 0.50, hence suggesting all three CDFs follow the same distribution at a minimum of 50\% significance level.
\section{Metallicity gradients along vertical and radial directions}\label{sec:Fe_Gradient_Rad_Vert}
\subsection{Vertical metallicity gradient}\label{sec:FeH_Z}
\begin{figure*}
\centering
\includegraphics[width=1\textwidth]{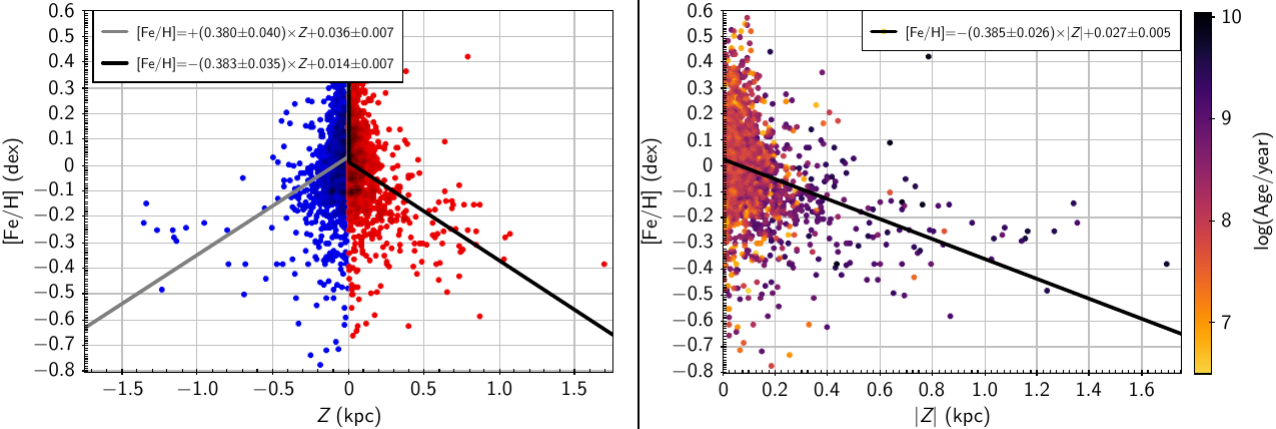}
\caption{Left-side panel: OC's metallicity as a function of their vertical distance (Z) from the Galactic plane. Linear fits for Z $<$ 0 and Z $>$ 0 are also shown as grey and black lines, respectively.
Right-side panel: OC's metallicity as a function of the magnitude of the vertical distance ($|Z|$) from the Galactic plane. A linear fit to the distribution is shown as the black straight line. The age of each cluster is also encoded in color as shown in the color bar.
}
\label{fig:FeH_Z}
\end{figure*}

The metallicity distribution in the Milky Way and its spatial variation is associated with the formation and evolution history of the Galaxy. The metallicity distribution at a particular point in the disk is linked with many parameters like the gas accretion rate, formation history, and evolution at that point of the disk. Previous studies, \cite{2005AstL...31..515M, 2006AstL...32..376M} and \cite{2008A&A...480...91S}, indicate that vertical metallicity distribution profiles can provide extremely meaningful ways for separating the thin disk from the thick disk. For our sample of OCs, metallicity as a function of vertical distance ($Z$) is shown in the left-side panel of Figure \ref{fig:FeH_Z}. The distribution readily suggests a decrease in [Fe/H] as we move away from the Galactic plane in both the Nothern and Southern hemispheres. To find the metallicity gradient in both the hemisphere we divide the sample about the centre of the Galactic mid-plane (i.e., at $Z$ = 0). Linear fits to cluster in the Southern (blue-colored points) and Northern hemispheres (red-colored points) are shown as grey and black lines, respectively, and obtained metallicity gradients are: 
\begin{equation}
\frac{d{\rm [Fe/H]}}{dZ} = 0.380 \pm 0.040\ {\rm dex\ kpc^{-1}}, ~~ -2 < Z < 0\ kpc
\label{eq:dFe_z_1}
\end{equation}
\begin{equation}
\frac{d{\rm [Fe/H]}}{dZ} = -0.383 \pm 0.035\ {\rm dex\ kpc^{-1}}, ~~~~ 0 < Z < 2 kpc
\label{eq:dFe_z_2}
\end{equation}
The magnitude of the metallicity gradient in both the Northern and Southern hemispheres are nearly the same and indicates that in both hemispheres metallicity changes at almost similar rates as we move away from the Galactic mid-plane. To find the average value of the metallicity gradient, as shown in the right-side panel of Figure \ref{fig:FeH_Z}, we plotted [Fe/H] as a function of absolute vertical distance from the Galactic plane. The metallicity gradient from a linear fit is:
\begin{equation}
\frac{d{\rm [Fe/H]}}{d|Z|} = -0.385 \pm 0.026\  {\rm dex\ kpc^{-1}}, ~~ |Z| <2\ kpc
\label{eq:dFe_mode_z}
\end{equation}
where the negative slope indicates that metallicity decreases as we move away from the Galactic mid-plane. These average vertical-metallicity gradients over a large distance are in agreement with the previous studies. For example, \cite{ChenHouWang2003AJ....125.1397C} found a vertical metallicity gradient of -0.295 $\pm$ 0.050 dex kpc$^{-1}$ using a sample of 118 OCs. Through a sample of 40,000 stars with low-resolution spectroscopy over 144 lines of sight, \cite{2014ApJ...791..112S} found a vertical metallicity gradient of $-0.243^{+0.039}_{-0.053}$ dex kpc$^{-1}$ in different [$\alpha$/Fe] subsamples. However, as readily evident from both panels in Figure \ref{fig:FeH_Z}, a single linear fit is insufficient to explain the full trend in metallicity as a function of vertical distance. For $|{\it Z}|\lessapprox 1$ kpc, the metallicity decreases rapidly while at the larger height, the change is relatively small.

\begin{figure}
\centering
\includegraphics[width=0.48\textwidth]{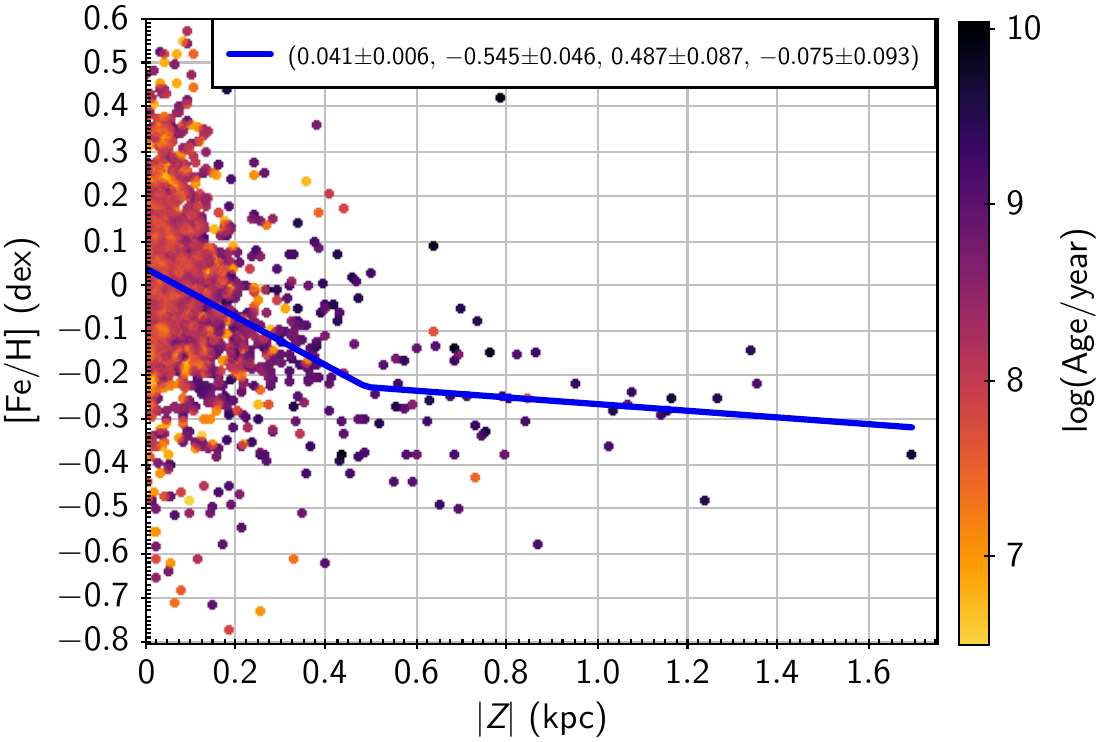}
\caption{Vertical metallicity distribution for sample clusters. The stepped linear fit to the data is shown as the blue line.
}
\label{fig:FeH_Z_stepped}
\end{figure}

To get a more accurate estimate for the vertical-metallicity gradient and also to find the vertical distance at which the radial-metallicity gradient changes significantly, we fitted the data with a combination of two linear regressions and the coefficients of the fitted functions are determined using iterative least square estimation following the procedure used in Section \ref{sec:FeH_logAge}. Based on the distribution in Figure \ref{fig:FeH_Z}, we assumed the initial values of b1, m1, C, and m2 as 1.0, 0.4, 1.0, and 0.0, respectively. The final fitted function is shown as the red line in Figure \ref{fig:FeH_Z_stepped}. The coefficients of the fitted functions are also provided in the legends of figure~\ref{fig:FeH_Z_stepped} in the form (b$_1$, m$_1$, C, m$_2$), where b$_1$ and m$_1$ are the y-axis intercept and slope for the first linear function, m$_2$ is the slope for the second linear function, and C is the point where these two function intersects. From the least square fitting, it is found that the two linear functions intersect at $|{\it Z}| = 0.487 \pm 0.087$ kpc and vertical-metallicity gradients are described as follows:
\begin{equation}
  \begin{array}{l}
    \frac{d{\rm [Fe/H]}}{d|Z|} = -0.545 \pm 0.046\ {\rm dex\ kpc^{-1}},\\~~~~~~~~~~~~~~~~~~~~~~~~~~ 0 < |Z| < 0.487 \pm 0.087\ {\rm kpc}
  \end{array}
\label{eq:dFe_z_stepped_1}
\end{equation}
\begin{equation}
  \begin{array}{l}
    \frac{d{\rm [Fe/H]}}{d|Z|} = -0.075 \pm 0.093\ {\rm dex\ kpc^{-1}},\\~~~~~~~~~~~~~~~~~~~~~~~  0.487 \pm 0.087 < |Z| \lessapprox 1.8\ {\rm kpc}
  \end{array}
\label{eq:dFe_z_stepped_2}
\end{equation}
This stepped vertical-metallicity gradient is in agreement with the currently accepted models of the Galaxy having a metal-rich disk (consisting of the thin and thick disk with scale heights of about 300 pc and 900 pc, respectively) and a metal-poor stellar halo (\cite[e.g.,][]{JustJahreiss2010MNRAS.402..461J, RixBovy2013A&ARv..21...61R, Matteucci2021A&ARv..29....5M} and references therein). In Figure \ref{fig:FeH_Z_stepped} (and also in the right-hand panel of Figure \ref{fig:FeH_Z}), cluster ages are also provided in the color of the data point. The figure readily suggests that the clusters at relatively larger vertical distances are comparatively old apart from being metal-poor. The lower metallicity in these clusters may be explained by their formation in the outer region of the Galactic disk at a relatively older time when the interstellar medium was relatively less enriched than the inner region of the Galactic disk. In Table \ref{table:FeH_Z_slope}, we summarize our results along with the previous reporting.
\subsection{Radial metallicity gradient}\label{sec:FeH_RGC}
The radial metallicity gradient is another important information to understand the chemical evolution of the Galactic disk and in turn the evolution of the Galaxy as the distribution of metallicity is not homogeneous across the Galaxy. It has been found that the metallicity in cluster population shows a decreasing trend with increasing distance from the GC (\cite{WuZhouMa2009MNRAS.399.2146W, 2010A&A...511A..56P, 2012AJ....144...95Y, 2015MNRAS.446.1411D, 2015A&A...580A..85M, CarreraBragagliaCantat-Gaudin2019AA...623A..80C, DonorFrinchaboyCunha2020AJ....159..199D, ZhongChenWu2020AA...640A.127Z, 2021ApJ...919...52Z, 2022AJ....164...85M, 2022Univ....8...87S, NetopilOralhanInc2022MNRAS.509..421N, MagriniViscasillasVazquezSpina2023AA...669A.119M}). The radial metallicity gradient and its evolution with age are among the most critical empirical constraints that one can put on the Galactic chemical evolution models. Most of these models show that the formation of clusters strongly influences the appearance and development of radial metallicity gradients (\cite{2001ApJ...554.1044C}) and the precise value of the metallicity gradient in the Galactic disk is an important parameter to constrain the chemical evolution models. The existence of such gradient across the Milky Way disk is well established through the observations of HII regions, disk stars, hot stars, star clusters, planetary nebula, Cepheid variables, and field stars (\cite{ChenHouWang2003AJ....125.1397C, 2010A&A...512A..19M}) and OCs (\cite{ CarreraBragagliaCantat-Gaudin2019AA...623A..80C, DonorFrinchaboyCunha2020AJ....159..199D, ZhongChenWu2020AA...640A.127Z, 2021ApJ...919...52Z, 2022AJ....164...85M, 2022Univ....8...87S, NetopilOralhanInc2022MNRAS.509..421N, MagriniViscasillasVazquezSpina2023AA...669A.119M}). An average gradient of about $-0.06$ dex kpc$^{-1}$ is observed in the Milky Way disk for most of the elements, e.g., O, S, Ne, Ar, and Fe. This magnitude of the observed gradients constrains the various parameters in the chemical evolution model, such as the time scales of star formation and infall (\cite{1995A&A...302...69P}) or any variations of the stellar initial mass function properties with metallicities (\cite{2001ApJ...554.1044C}).

\begin{figure}
\centering
\includegraphics[width=0.48\textwidth]{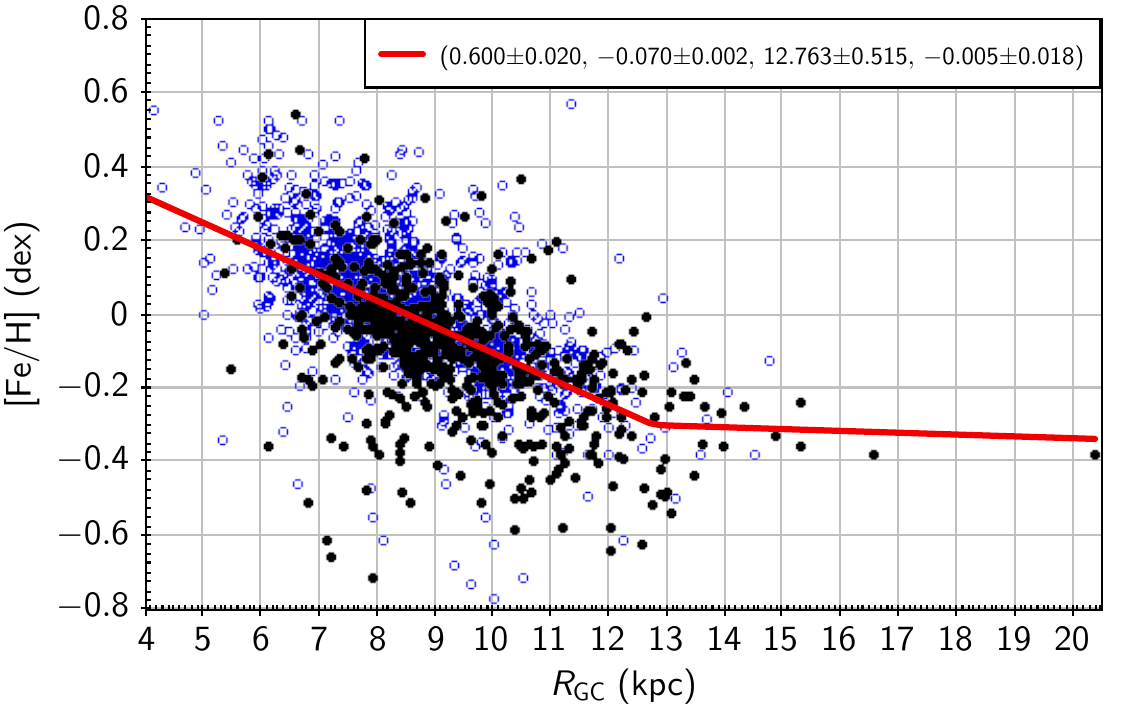}
\caption{Metallicity as a function of radial distance from the Galactic centre ($R_{\rm GC}$) for the sample of clusters. Open blue circles and filled black circles are the clusters from sample OCP and OCS, respectively. The red line shows the fitted stepped linear regression to the total sample, coefficients (along with corresponding standard errors) for which are provided in the legends in the form (b$_1$, m$_1$, C, m$_2$), where b$_1$ and m$_1$ are the y-axis intercept and slope for the first linear function, m$_2$ is the slope for the second linear function, and C is the point where these two function intersects.
}
\label{fig:FeH_Rgc}
\end{figure}

Star clusters are considered one of the most important celestial sources for investigating the metallicity gradient along the Galactic disk as their distance and age are derived very precisely and are available in a wide range. Metallicity as a function of radial distance from the GC ($R_{\rm GC}$) for our sample OC is shown in Figure \ref{fig:FeH_Rgc}. On average, as expected, the figure suggests a decreasing trend in [Fe/H] with an increase in distance from the GC. Additionally, the figure also suggests that the decrease in [Fe/H] with an increase in $R_{\rm GC}$ is not a simple linear function but at least a combination of two linear functions.

To find radial-metallicity gradients and the radial distance at which the radial-metallicity gradient changes, we fitted the data with a combination of two linear functions, and the coefficients of the fitted functions are determined using iterative least square estimation adopting the procedure followed in Section \ref{sec:FeH_logAge}. Based on the distribution in Figure \ref{fig:FeH_Rgc}, we assumed the initial values of b1, m1, C, and m2 as 1.0, -0.05, 12.0, and -0.03, respectively.
The final fitted function is shown as the red line in the figure. The coefficients of the fitted functions are also provided in the legends of the figure in the form (b$_1$, m$_1$, C, m$_2$), where b$_1$ and m$_1$ are the y-axis intercept and slope for the first linear function, m$_2$ is the slope for the second linear function, and C is the point where these two function intersects. The two linear functions intersect at $R_{\rm GC}$ of $12.763 \pm 0.515$ kpc and gradients in $R_{\rm GC}$ - metallicity distributions are found as:
\begin{equation}
  \begin{array}{l}
    \frac{d{\rm [Fe/H]}}{d {R_{\rm GC}}} = -0.070 \pm 0.002\  {\rm dex\ kpc^{-1}},\\~~~~~~~~~~~~~~~~~~~~ 4.0 \lesssim R_{\rm GC} \leq 12.763 \pm 0.515\ {\rm kpc}
  \end{array}
\label{eq:dFe_Rgc_1}
\end{equation}
\begin{equation}
  \begin{array}{l}
\frac{d{\rm [Fe/H]}}{d {R_{\rm GC}}} = -0.005 \pm 0.018\  {\rm dex\ kpc^{-1}},\\~~~~~~~~~~~~~~~~~~ 12.763 \pm 0.515 < R_{\rm GC} \lesssim 20.5 \ {\rm kpc}
  \end{array}
\label{eq:dFe_Rgc_2}
\end{equation}
The existence of the two-step linear distribution can be explained in most evolution models by assuming different infall and star formation rates for the inner and outer disks. A similar two-step distribution was also noticed by \cite{2011MNRAS.417..698L, GozhaBorkovaMarsakov2012AstL...38..506G, 2022AJ....164...85M, MagriniViscasillasVazquezSpina2023AA...669A.119M}, and others. All these studies have found a discontinuity in the radial metallicity gradient at $R_{\rm GC} \sim 10-12$ kpc with a steeper gradient in the inner disk region, and a flatter gradient or a plateau in the outer disk region. However, some other studies have not seen such two-step distribution although they found a decreasing trend in metallicity with increasing $R_{\rm GC}$ e.g. \cite{FrielJanesTavarez2002AJ....124.2693F, ChenHouWang2003AJ....125.1397C, 2009A&A...494...95M, GaiaCollaborationRecioBlancoKordopatis2023AA...674A..38G}, etc.

Our estimated radial metallicity gradients are in close agreement with some of the recent determinations of metallicity gradients derived using samples of OCs. A comparison of radial metallicity gradients from some of the recent studies based on OCs along with our estimates is provided in Table \ref{table:FeH_Rgc_slope}. Most of these studies suggest a radial metallicity gradient of about -0.06 dex~kpc$^{-1}$. The radial metallicity gradient provides vital information on radial migration which plays an important role in the redistribution of stellar populations, particularly the older ones, in our Galaxy. It is believed that radial migration in OCs may be the reason for the flattening of the radial metallicity gradient over a period of time (\cite{2021ApJ...919...52Z, 2022A&A...660A.135V}). It is believed that there is a deficiency of low-metallicity clusters in the inner disk migrating from the outer disk, as the chance of survival in the high Galactic potential of the inner disk is low. On the other hand, clusters from the more metal-rich inner Galactic disk can migrate farther into the outer disk where the potentials of the spiral arm and bar are weaker, resulting in the enhancement of the mean metallicity of the outer disk. As a consequence, the radial metallicity gradient is steeper in the inner disk while flattening out towards the large galactocentric distance. Various earlier studies using different tracers such as planetary nebulae, classical Cepheids, globular clusters etc., also suggested that the radial metallicity gradient becomes slightly flatter with time (\cite[e.g.][]{FrielJanesTavarez2002AJ....124.2693F, ChenHouWang2003AJ....125.1397C, 2009IAUS..254P..38M, 2011AJ....142...51L, GenovaliLemasleBono2014AA...566A..37G, 2023A&A...678A.195D}). It was contemplated by \cite{2014ApJ...788...89T} that the radial metallicity gradient was positive at the time of formation of the thick disk which subsequently became negative during the transition phase of disk formation from thick to thin disk. It became flatter by the time of the formation of the thin disk. They credited this evolution of disk to the gas in-fall history having a shorter time scale in the inner disk and a relatively longer time scale in the outer disk, which is often called the 'inside-out' scenario in the disk formation.
\subsection{Age dependence of radial and vertical metallicity gradients} \label{sec:Fe_Gradient_Rad_Vert_ageBinned}
One of the crucial questions in the chemical evolution of the Galaxy is how the metallicity gradients have evolved over the last few Gyrs. As the overall metallicity gradient may introduce a bias due to the mix of different age OCs, we may need to restrict the sample to OCs into different age bins in order to understand the evolution in the metallicity gradients along the radial and vertical directions with time. The age dependence of metallicity gradient has been investigated in the past using a variety of sources (\cite[e.g.,][and references therein]{2021ApJ...922..189V}). We therefore split our sample broadly into three age bins, including the very young age bin ($<$ 20 Myr), the young to intermediate age bin (20 Myr - 700 Myr), and the old age bin ($>$ 700 Myr). Since we have less than 10\% of the OCs older than 1 Gyr, we have not split bins in the older age regime. Table \ref{table:FeH_Rgc_Z_slopes} shows the slopes for the graphs in various age bins. Along the radial direction, the three age populations have almost similar metallicity gradients except for the youngest clusters which have a slightly shallower gradient than the intermediate-age clusters. The lower (or flatter) gradient in the case of the older population is in agreement with previous studies and models and could be explained by the chemical evolution in the Galactic disk (\cite{2002ChJAA...2..226C, 2016A&A...591A..37J, ZhongChenWu2020AA...640A.127Z}) and radial migration (\cite{2017A&A...600A..70A, NetopilPaunzenHeiter2016AA...585A.150N}). For example, in the MCM model, radial migration is expected to flatten the radial metallicity gradient for clusters older than one Gyr (\cite{2014A&A...572A..92M}).

For the three age distributions, we also examined the vertical metallicity gradient i.e. change in the metallicity as a function of Galactic disk thickness. We found a steeper slope for the young and intermediate-age OCs while it is shallower for the old OCs. The estimated slope values are given in the third column of Table \ref{table:FeH_Rgc_Z_slopes}. This behavior of OCs is well expected because most of the young and intermediate-age OCs lie closer to the metal-rich thin Galactic disk while older OCs are located farther away in the metal-poor outer disk. \cite{CarreraBragagliaCantat-Gaudin2019AA...623A..80C}, however, do not find any evidence of the presence of a vertical metallicity gradient, at least above the 1-$\sigma$ level. A further examination of the vertical evolution of the metallicity gradient is performed in the next section.
\subsection{Vertical evolution of radial metallicity gradient}
\begin{figure}
\begin{center}
\includegraphics[width=0.48\textwidth]{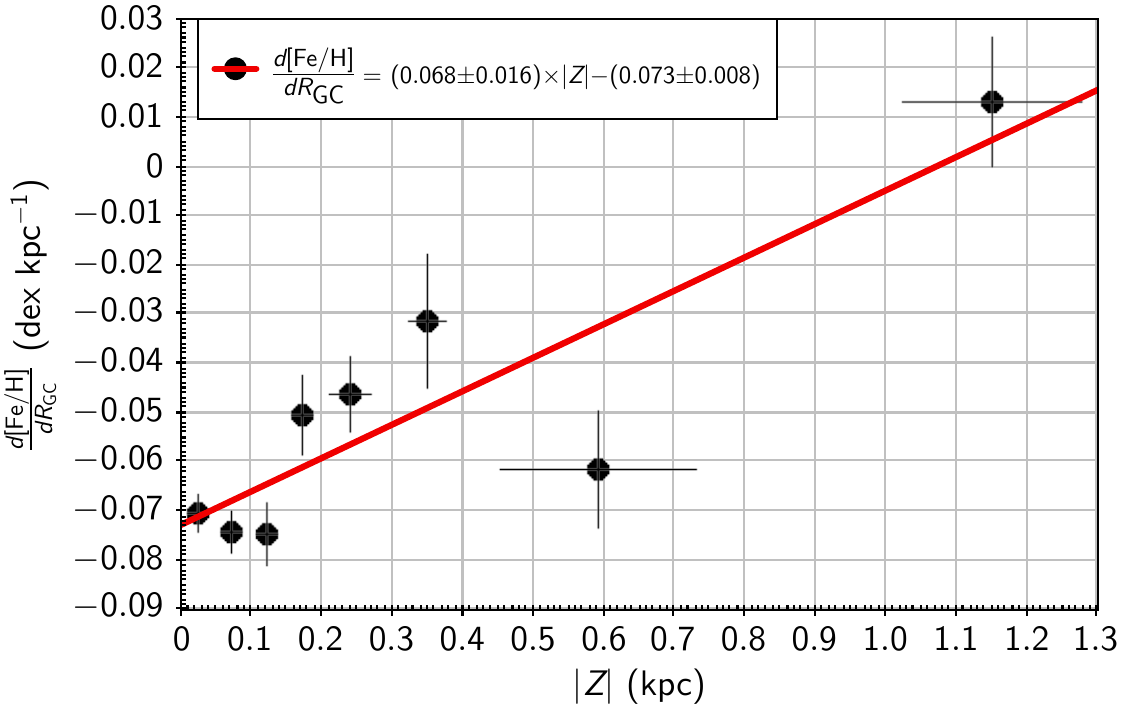}
\caption{Variation in $\frac{d{\rm [Fe/H]}}{dR_{\rm GC}}$ as a function of absolute vertical distance ($|Z|$) for the the sample of OCs. Uncertainties are shown as black cross bars. The linear fit to the data is shown as a red color line.}
\label{fig:FeH_RGC_vs_Z}
\end{center}
\end{figure}

The study of the relation between the metallicity and the location of the cluster on the Galactic disk is an important tool for the study of the structure formation and evolution of the Galaxy (\cite[e.g.,][]{ ZhongChenWu2020AA...640A.127Z}). We also investigated the evolution of radial metallicity gradients in the vertical direction of the Galactic plane as clusters are widely distributed in the vertical direction of the Galactic disk. The effect of scale height on the rate of change in metallicity variation with the $R_{\rm GC}$ has been analyzed by plotting the slope of the radial metallicity gradients as a function of the absolute value of $Z$ which is shown in Figure ~\ref{fig:FeH_RGC_vs_Z}. $|Z|$ for our sample ranges from 0 to about 1.8 kpc (excluding one lone cluster located at about 2.6 kpc). While most of the clusters are located near the Galactic plane, there are fewer clusters at larger vertical distances. Therefore, we considered a varying bin size in the $Z$ scale. We considered a bin width of 50 pc for $|Z| < 200$, 100 pc for $200 < |Z| < 400$, and 500 pc for $|Z| > 400$ by making sure that there are enough clusters in the selected bins to get a proper estimate of radial metallicity gradient. As shown in Figure \ref{fig:FeH_RGC_vs_Z}, the variation of $\frac{d{\rm [Fe/H]}}{dR_{\rm GC}}$ as a function of absolute vertical distance  $|Z|$ for our sample of OCs follows an increasing trend with an increase in $|Z|$. Estimated standard errors in $\frac{d{\rm [Fe/H]}}{dR_{\rm GC}}$ and mean $|{\it Z}|$ are shown as black coloured cross bars. At larger vertical distances, the estimated errors are larger because of the smaller sample size in corresponding vertical distance bins. A linear fit to the distribution is shown as a thick red line and corresponding coefficients are shown in the legends. The radial metallicity gradient ($\frac{d{\rm [Fe/H]}}{dR_{\rm GC}}$) as a function of vertical distance from the Galactic plan is found to vary at a rate of $0.068 \pm 0.016$ dex~kpc$^{-1}$~kpc$^{-1}$, suggesting a higher rate of change of metallicity with $R_{\rm GC}$ at larger vertical distance from the Galactic plane. The radial metallicity gradient at the centre of the Galatic plane is estimated as $-0.073 \pm 0.008$ dex~kpc$^{-1}$ which is in agreement with our previous estimate in Section \ref{sec:FeH_RGC}. 
\section{Discussion and Conclusions}\label{sec:discuss}
In this study, we used the largest sample of 1879 open clusters to understand the distribution and evolution of metallicity in the Galactic disk. The cluster sample was compiled from the literature with available metallicity information along with other information like age, position coordinates, distances, and radial and vertical distances. About 90\% of the OCs in our sample are younger than 1 Gyr with the oldest being about 10 Gyr old. Radially and vertically, about 90\% of clusters in our sample are within a heliocentric distance of 3 kpc while about 97\% of clusters are within a vertical distance of 500 pc, practically restricting our study to the Galactic disk.

Age-metallicity relation provides an important constraint on the theoretical models of the disk and thus has been studied multiple times in the past. The study of metallicity evolution for our sample of OCs did not find a strict age-metallicity relation but a stepped linear evolution of metallicity in the Galaxy was observed with a discontinuity at $\log ({\rm age/year}) = 8.378 \pm 0.093$ at the age of about 240 Myr. OCs older than 240 Myr follow a decreasing trend in metallicity with an increase in age with an age-metallicity gradient of $-0.031\pm0.006$ dex/Gyr which is in agreement with some of the recent studies as well as Galactic evolutionary models. The slightly higher average metallicity in the intermediate age clusters compared to the average metallicity in the young ones agrees with findings in earlier studies (\cite{2010A&A...511A..56P, ZhongChenWu2020AA...640A.127Z}). Interestingly, the sample of OCs younger than about 240 Myr follows a slightly increasing trend in metallicity with an increase in age. The radial and vertical migration of young OCs in the disk is suspected to be one of the main reasons for this weak correlation between $\log \rm (age)$ and [Fe/H] for younger clusters. However, no strong correlation has been found to draw any meaningful conclusion. Despite a large scatter in the age-metallicity relation in our study, it is crucial to observe the slightly different age-metallicity relation for two different samples of clusters which possibly put distinct formation constraints on the Galactic thin and thick disc in the modeling the Milky Way.

It is well understood that the metallicity in the inner region of the Galactic disk is increasing with time (\cite[e.g.,][and references therein]{ReddyTomkinLambert2003MNRAS.340..304R,HaywoodDiMatteoLehnert2013A&A...560A.109H}). Hence, the younger clusters with lower metallicity must have either formed away from the Galactic plane or in the anti-GC direction. To see whether later is the reason behind lower metallicity in young clusters, we looked into the distribution of metallicity in the Galactic plane by plotting metallicity as a function of Galactic longitude. The OCs in the anti-GC direction do have lower metallicity compared to the OCs in the GC direction, possibly owing to the differences in timelines of gas in-falls and formation of clusters in the GC and anti-GC direction. Our samples of YOCs, IOCs, and OOCs are found to equally populate both the GC and anti-GC directions, hence leaving vertical migration as one of the likely reasons for slightly lower metallicity in young clusters.

Using our sample of clusters, we further explored the vertical and radial metallicity gradients in the Galactic disk. Metallicity was found to follow a stepped variation with vertical distance from the Galactic plane. Near the Galactic plane, with $|Z| < 0.487 \pm 0.087$ kpc, we estimated the vertical metallicity gradient of $-0.545 \pm 0.046$ dex kpc$^{-1}$, while for large vertical distance having $0.487 \pm 0.087 < |Z| \lesssim 1.8$ kpc, we found a lower vertical metallicity gradient of $-0.075 \pm 0.093$ dex kpc$^{-1}$. The lower metallicity gradient at large vertical distances compared to one at smaller vertical distances agrees with the Galactic chemical evolution models. We found that most of the OCs at large vertical distances are older compared to the majority of the clusters located near the Galactic plane. This difference in the ages of clusters from the two vertical regions is believed to be the main reason for the flatter vertical metallicity gradient at large vertical distances compared to the steep vertical metallicity gradient at smaller vertical distances.

Similar to the vertical direction, the change in metallicity in the radial direction is also found to follow a stepped linear relation. For a radial distance between about 4.0 to 12.8 kpc, we found a radial metallicity gradient ($\frac{d{\rm [Fe/H]}}{dR_{\rm GC}}$) of $-0.070 \pm 0.002$ dex kpc$^{-1}$ while for radial distance between about 12.8 to 20.5 kpc, we found a much smaller radial metallicity gradient of $-0.005 \pm 0.018$ dex kpc$^{-1}$. Thus the OCs in the outer Galactic disc are generally more metal-poor than the OCs in the inner Galactic disc and in the solar neighborhood. Although a shallower metallicity gradient in the region 12.8-20.5 kpc may be biased due to the relatively smaller number of OCs at larger Galactocentric distances but it could also be the result of radial migration of clusters in the Galactic disk (\cite{2021ApJ...919...52Z}). Using a smaller sample of 295 OCs within a Galactocentric distance of 7-15 kpc, \cite{ZhongChenWu2020AA...640A.127Z} reported a steeper slope of -0.252$\pm$0.039 dex~kpc$^{-1}$. It should also be noted that a significant variation in the slope as well as the turn-off point in the radial metallicity gradient among different studies, comes from the choice of the cluster sample, selected range of $R_{GC}$, and unequal vertical heights. Overall, our radial metallicity gradient estimates agree with most of the recent studies (\cite{2020JApA...41...38R, 2021ApJ...919...52Z, 2022AJ....164...85M}). 

One of the key questions in Galactic chemical evolution models is the evolution of the radial metallicity gradients over time and the answer is not settled yet. We therefore examined the time evolution of the metallicity gradients, both in radial and vertical directions, with age by dividing, the clusters into three age bins of $<$ 20 Myr, 20-700 Myr, and $>$ 700 Myr. We observed that these gradients are shallower for the oldest age bin while not much difference was noticeable in young and intermediate age clusters. The time evolution of abundance gradients has also been examined in the past but an unequivocal result has not been found so far. While \cite{2018MNRAS.480L..38V, 2018MNRAS.481.1645M} suggested a flatter metallicity gradient with time, there are a few studies like \cite{2001ApJ...554.1044C, 2013MNRAS.435.2918M}  which suggested a steepening in gradient over time. However, variation is only prominent over a longer time scale, and the limited temporal coverage of the present cluster sample, where only a small number of OCs are available beyond the 1 Gyr period, in no way sheds any more light on this discussion. We refer \cite{MagriniViscasillasVazquezSpina2023AA...669A.119M} for a more detailed discussion on the temporal evolution of the metallicity gradients.

We further studied the variation of radial metallicity gradient with distance from the Galactic plane and found that the radial metallicity gradient linearly increases with an increase in vertical distance and obtained a radial metallicity gradient slope of $\frac{d{\rm [Fe/H]}}{dR_{\rm GC}} =  0.068 \pm 0.016$ dex~kpc$^{-1}$~kpc$^{-1}$ as a function of vertical distance from the Galactic plane. This agrees with the Galactic evolutionary models, for example, see \cite{2014ApJ...788...89T} and references therein. In the case of a thin disk, which has a scale height of about 300 pc, the radial metallicity gradient is highly negative even though linearly increasing with vertical distance from the Galactic plane. However, for the thick disk (having a typical scale height of about 900 pc), the radial metallicity gradient is slightly high and approaches zero at about 1 kpc. \cite{2014ApJ...788...89T} suggested that the radial metallicity gradient was positive at the time of formation of the thick disk but subsequently became negative during the transition phase of the disk formation from thick to thin disk. The gradient became flatter by the time of the formation of the thin disk. This change in radial metallicity gradient with vertical distance is believed to be related to the gas-infall history in the Galaxy. A large negative radial metallicity gradient near the Galactic plane (i.e., in the thin disk) but a higher gradient in the case of the thick disk (i.e., large vertical distance) can be explained by shorter and longer time scales, respectively, for the gas in-fall.

\section*{Author contributions}
YJ: writing original draft, conceptualization, methodology, and resources. D: formal analysis, writing–review and editing, data investigation, and validation. SM: data curation.

\section*{Acknowledgments}
We thank both the referees for their constructive and insightful suggestions, which significantly improved the quality of the paper. We also thank Vaibhav Pant for his help in computing the data. YCJ thanks master students Asish Philip and Ananya Bandopadhyay, who contributed to this project as a part of their summer project internship.

\bibliographystyle{Frontiers-Harvard}
\bibliography{main}
\begin{table*}
\renewcommand{\thetable}{\arabic{table}}
\centering
\caption{The final catalogue of 1879 OCs used in this study. The entries in the spectral resolution column (Resol.) are left blank for the studies where adopted metallicities are based on photometric data.}
\tiny
\label{table:Data_Table}
\begin{tabular}{rlrrrrrrccccl}
\hline
S.N. & Cluster ID & RA & DEC & $X$ & $Y$ & $Z$ & $R_{GC}$ & $\log(age)$ & [Fe/H] & e[Fe/H] & Resol. & Reference \\
\hline
   1    &        ASCC 10  &  51.807  &  34.945  &  -525.95  &   237.45  &  -185.60  &   8.681 &  7.90  & -0.024 &   0.018 &   1800    &\cite{FuBragagliaLiu2022AA...668A...4F}\\
   2    &       ASCC 101  & 288.408  &  36.377  &   145.43  &   360.67  &    79.86  &   8.013 &  8.10  &  0.004 &   0.008 &           &\cite{DiasMonteiroMoitinho2021MNRAS.504..356D}\\
   3    &       ASCC 103  & 294.031  &  35.735  &   170.63  &   457.26  &    62.16  &   7.993 &  7.90  &  0.115 &   0.024 &   1800    &\cite{FuBragagliaLiu2022AA...668A...4F}\\
   4    &       ASCC 105  & 295.540  &  27.402  &   235.80  &   459.86  &    18.83  &   7.928 &  7.99  &  0.046 &   0.024 &   1800    &\cite{FuBragagliaLiu2022AA...668A...4F}\\
   5    &       ASCC 106  & 295.286  &   1.494  &   503.30  &   422.53  &  -119.95  &   7.659 &  8.06  &  0.029 &   0.005 &           &\cite{DiasMonteiroMoitinho2021MNRAS.504..356D}\\
   6    &       ASCC 107  & 297.164  &  21.994  &   445.92  &   739.48  &   -28.62  &   7.740 &  7.05  &  0.353 &   0.013 &           &\cite{DiasMonteiroMoitinho2021MNRAS.504..356D}\\
   7    &       ASCC 108  & 298.355  &  39.328  &   286.80  &  1025.76  &   112.52  &   7.931 &  7.91  & -0.106 &   0.067 &   1800    &\cite{FuBragagliaLiu2022AA...668A...4F}\\
   8    &        ASCC 11  &  53.029  &  44.877  &  -729.52  &   412.45  &  -135.99  &   8.890 &  8.45  & -0.360 &   0.015 &  22500    &\cite{MyersDonorSpoo2022AJ....164...85M}\\
   9    &       ASCC 110  & 300.772  &  33.549  &   529.77  &  1491.24  &    37.78  &   7.765 &  8.79  &  0.140 &   0.024 &           &\cite{DiasMonteiroMoitinho2021MNRAS.504..356D}\\
  10    &       ASCC 111  & 302.960  &  37.544  &   215.04  &   789.73  &    28.95  &   7.974 &  7.90  &  0.080 &   0.008 &           &\cite{DiasMonteiroMoitinho2021MNRAS.504..356D}\\
.. &... & ... & ... & ... & ... & ... & ... & ... & ... & ... & ... & \\
.. &... & ... & ... & ... & ... & ... & ... & ... & ... & ... & ... & \\
1879    &     vdBergh 92  & 106.186  & -11.333  &  -793.59  &  -780.59  &   -43.40  &   8.978 &  6.75  &  0.025 &   0.007 &           &\cite{DiasMonteiroMoitinho2021MNRAS.504..356D}\\

\hline
\end{tabular}\\
\small
\noindent\footnotesize{The entire table is available in the online version in a machine-readable format.}
\end{table*}
%
%

\begin{table}
\centering
\caption{Comparison of age-metallicity slope among different studies based on open clusters. The number of clusters ($N$) used in each study is provided in the second column.}
\label{table:Age_FeH_slope}
\begin{tabular}{@{}c c l@{}}
\hline
Slope             & $N$ & Reference\\
(dex Gyr$^{-1}$)  &     & \\                   \hline
-0.026            & 57    & \cite{2010AA...511A..56P} \\
$-0.022\pm0.0008$ & 295   & \cite{ZhongChenWu2020AA...640A.127Z} \\
$-0.031\pm0.006$  & 786   & This work\\
\hline
\end{tabular}
\end{table}

\begin{table}
\centering
\caption{Comparison of vertical metallicity gradient $\left(\frac{d{\rm [Fe/H]}}{d |Z|}\right)$ reported in the previous studies with estimates in this work. The number of clusters ($N$) used in each study is provided in the third column.}
\label{table:FeH_Z_slope}
\begin{tabular}{cccl}
\hline
$\frac{d{\rm [Fe/H]}}{d|Z|}$ & $|Z|$ & $N$ & Reference \\
(dex kpc$^{-1}$) & (kpc) & & \\  \hline
$-0.34\pm0.03$ & $< 1.30$  & 63  & \cite{PiattiClariaAbadi1995AJ....110.2813P} \\
$-0.295\pm0.050$ & $< 1.40$   & 118  & \cite{ChenHouWang2003AJ....125.1397C} \\
$-0.252\pm0.039$ & $< 0.90$   & 183  & \cite{ZhongChenWu2020AA...640A.127Z} \\
$-0.545\pm0.046$ & $< 0.487$  & 1814 & This work \\
$-0.075\pm0.093$ & 0.487-1.80 &   58 & This work \\  
\hline
\end{tabular}
\end{table}
%
\begin{table}
\centering
\caption{Comparison of radial metallicity gradient $\left(\frac{d{\rm [Fe/H]}}{d {R_{\rm GC}}}\right)$ among different studies based on the sample of open clusters. The number of clusters ($N$) used in each of the studies is provided in the third column.}
\label{table:FeH_Rgc_slope}
\begin{tabular}{@{}c c c l@{}}
\hline
$\frac{d{\rm [Fe/H]}}{d {R_{\rm GC}}}$ & $R_{\rm GC}$ & $N$ & Reference\\
(dex kpc$^{-1}$) & (kpc)       &      & \\ 
\hline
$-0.059\pm0.010$ & 7-16        & 39   & \cite{FrielJanesTavarez2002AJ....124.2693F} \\
$-0.063\pm0.008$ & $<$17       & 118  & \cite{ChenHouWang2003AJ....125.1397C} \\
$-0.056\pm0.007$ & $<$17       & 488  & \cite{WuZhouMa2009MNRAS.399.2146W} \\
$-0.051 \pm 0.003$ & 5-15      & 127  & \cite{GenovaliLemasleBono2014AA...566A..37G} \\
$-0.061\pm0.004$ & 7-13        & 19   & \cite{DonorFrinchaboyCunha2018AJ....156..142D} \\
$-0.052\pm0.003$ & 6-13        & 46   & \cite{CarreraBragagliaCantat-Gaudin2019AA...623A..80C} \\
$-0.077\pm0.007$ & 6-14.5      & 90   & \cite{CarreraBragagliaCantat-Gaudin2019AA...623A..80C}  \\
$-0.068\pm0.001$ & 6-13.9      & 71   & \cite{DonorFrinchaboyCunha2020AJ....159..199D} \\
$-0.053\pm0.004$ & 7-15        & 295  & \cite{ZhongChenWu2020AA...640A.127Z} \\
$-0.074\pm0.007$ & 6-20        & 225  & \cite{2021ApJ...919...52Z} \\
$-0.066\pm0.006$ & 6-15.5      & 157  & \cite{2021ApJ...919...52Z} \\
$-0.076\pm0.009$ & 6-16.5      & 134  & \cite{SpinaTingDeSilva2021MNRAS.503.3279S} \\
$-0.073\pm0.002$ & 6-11.5      & 94   & \cite{2022AJ....164...85M}  \\
$-0.032\pm0.002$ & 11.5-16.0   & 56   & \cite{2022AJ....164...85M} \\
$-0.054\pm0.008$ & 5-12        & 503  & \cite{GaiaCollaborationRecioBlancoKordopatis2023AA...674A..38G} \\
$-0.064\pm0.007$ & 5-24        & 175  & \cite{2022Univ....8...87S} \\
$-0.058$         & 6-21        & 136  & \cite{NetopilOralhanInc2022MNRAS.509..421N} \\
$-0.054\pm0.004$ & 6-21        &  62  &  \cite{MagriniViscasillasVazquezSpina2023AA...669A.119M} \\
$-0.081\pm0.008$ & 6-11.2      &  42  & \cite{MagriniViscasillasVazquezSpina2023AA...669A.119M} \\
$-0.044\pm0.014$ & 11.2-21     &  20  & \cite{MagriniViscasillasVazquezSpina2023AA...669A.119M} \\
$-0.070\pm0.002$ & 4.0-12.8    & 1837 & This work \\
$-0.005\pm0.018$ & 12.8-20.5   & 35   & This work \\
\hline
\end{tabular}
\end{table}
%
\begin{table}
\renewcommand{\thetable}{\arabic{table}}
\centering
\caption{Radial and vertical metallicity gradients for OCs of different age groups. The number of clusters ($N$) in each of the age bin are provided in the fourth column.
}
\label{table:FeH_Rgc_Z_slopes}
\begin{tabular}{cccc}
\hline
Age & $\frac{d{\rm [Fe/H]}}{d {R_{\rm GC}}}$ & $\frac{d{\rm [Fe/H]}}{dZ}$ & $N$\\
(Myr) & (dex~kpc$^{-1}$) & (dex~kpc$^{-1}$) & \\ \hline
$<$20   & -0.063 $\pm$ 0.005 &  -0.427 $\pm$ 0.148  &  410\\
20-700  & -0.071 $\pm$ 0.003 &  -0.459 $\pm$ 0.061 &  1114\\
$>$700   & -0.058 $\pm$ 0.004 &  -0.245 $\pm$ 0.032  &  349\\ 
\hline
\end{tabular}
\end{table}
%
%
%
%
%
%
%
%
%
%

\end{document}